# The pseudorapidity distributions of the produced charged particles in *p-p* collisions at center-of-mass energies from 23.6 to 900 GeV

Z. J. Jiang[*] and H. L. Zhang

College of Science, University of Shanghai for Science and Technology, Shanghai 200093, China

**Abstract.** In *p-p* collisions, there are two leading particles. One is in projectile fragmentation region. The other is in target fragmentation region. The investigations of present paper show that, just like in nucleus-nucleus collisions, the revised Landau hydrodynamic model alone is not enough to have a good description to the measured pseudorapidity distributions of the produced charged particles in *p-p* collisions. Only after the leading particles are taken into account as well, can the experimental data be matched up properly. The theoretical model works well in *p-p* collisions in the whole available energy region from $\sqrt{s}$ =23.6 to 900 GeV.

**1.** Relativistic hydrodynamics provides us a theoretical framework for describing the motion of a continuous flowing medium. It is now widely used to depict various processes for system large to the whole universe and small to the matter created in high energy hadronic or nuclear collisions. The experimental observations, such as the elliptic flow, the single-particle spectra, and the two-particle correlation functions for the matter produced in collisions have indeed shown the existence of a collective effect similar to an almost perfect fluid motion [1-4], and can be reasonably well reproduced by hydrodynamic approach. This gives us a confidence to believe that the relativistic hydrodynamics might be one of the best tools for the description of the space-time evolution of the matter generated in collisions. Hence, in recent years, the relativistic hydrodynamics has become one of the most active research areas, and has got more and more experimental approvals [5-19].

One of the important applications of the hydrodynamic model is the analysis of the pseudorapidity distributions of the produced charged particles in hadron or heavy ion collisions. A wealth of such distributions has been accumulated in experiments [20-25]. In our previous work [6], by taking into account the contributions from leading particles, we have once successfully used the revised Landau hydrodynamic model in describing such experimental measurements in nucleus-nucleus collisions at energies of RHIC (Relativistic Heavy Ion Collider) at BNL (Brookhaven National Laboratory). Now, what we are concerned is that whether the model can still work in hadron, such as in *p-p* collisions. To clarify this problem is just the subject of this paper. We can see that, just as in nucleus-nucleus collisions, the total contributions from both revised Landau hydrodynamic model and leading particles are in good accordance with experimental data measured in *p-p* collisions at energies from $\sqrt{s}$ =23.6 to 900 GeV [20, 21].

---

[*]E-mail address: Jzj265@163.com (Zhijin Jiang)

**2.** The revised Landau hydrodynamic model is based on the following assumptions.

**(1)** The matter with high temperature and high density created in collisions is taken as a massless perfect fluid, which meets the equation of state

$$\varepsilon = 3P, \quad (1)$$

where $\varepsilon$ is the energy density, and $P$ is the pressure. This assumption is now well favored by experimental observations [1-4]. The investigations of the lattice gauge field theory have also shown that the above relation is approximately relevant for the matter with temperature $T > 240$ MeV [26, 27].

**(2)** During the process of expansion, the fluid quickly achieves local thermal equilibrium. The expansion is adiabatic, and the number of the produced charged particles is proportional to entropy [28, 29]. This means that the entropy in each fluid element or in the whole fluid body is conserved during the hydrodynamic evolution, and the total number of the observed particles can be determined from the initial entropy of the system.

**(3)** The expansion of fluid undergoes the following two stages [28, 29]. Stage 1: During the fast longitudinal expansion along colliding direction (taken as *z* axis), there is a simultaneous slow transverse expansion, and the expansions in these two directions advance independently. Stage 2: As the transverse displacement of a fluid element arrives at the initial transverse dimension of the overlap region, the pressure in this fluid element may be neglected. Its rapidity is frozen and therefore remains unchanged. It will have a conic flight with a certain polar angle. The rapidity of the observed particles is determined by that of the fluid element at freeze-out time.

According to assumption (1), we can get the expansion equation of the fluid along longitudinal *z* direction as

$$\frac{\partial \varepsilon}{\partial t_+} + 2\frac{\partial \left(\varepsilon e^{-2y}\right)}{\partial t_-} = 0,$$
$$2\frac{\partial \left(\varepsilon e^{2y}\right)}{\partial t_+} + \frac{\partial \varepsilon}{\partial t_-} = 0,$$
(2)

where $y$ is the rapidity of the fluid element, and

$$t_+ = t + z,$$
$$t_- = t - z,$$

are the light-cone coordinates. The solution of Eq. (2) is

$$\varepsilon(y_+, y_-) = \varepsilon_0 \exp\left[-\frac{4}{3}\left(y_+ + y_- - \sqrt{y_+ y_-}\right)\right],$$
(3)

where

$$y_\pm = \ln\left(\frac{\tau}{\Delta} e^{\pm y}\right),$$

$\tau$ is the proper time, and

$$\Delta = \frac{\sqrt{d^2 - b^2}}{\gamma}$$

is the thickness of the overlap region along $z$ direction for two protons with diameter $d$ colliding at impact parameter $b$. $\gamma = \sqrt{s}/2m_p$ is the Lorentz contract factor, $\sqrt{s}/2$ is the center-of-mass energy per proton, and $m_p$ is its mass.

For slow transverse expansion, it follows the equation

$$\frac{4}{3}\varepsilon \cosh^2 y \frac{\partial v_\phi}{\partial t} = -\frac{\partial P}{\partial \rho},$$
(4)

where $v_\phi$ is the transverse 3-velocity in the direction with the azimuthal angle $\phi$, and $\rho$ is the transverse displacement in this direction. The solution of above equation is

$$\rho(t) = \frac{t^2}{4d_\phi \cosh^2 y},$$
(5)

where $d_\phi$ is the initial distance between the two corresponding points on the boundary of the overlap region at azimuthal angle $\phi$.

Furthermore, by using the solutions of Eqs. (3) and (5), we can get the rapidity distribution of the produced charged particles in accordance with assumptions (2) and (3)

$$\frac{\mathrm{d}^2 N_{\text{Fluid}}(y,\sqrt{s},\phi)}{\mathrm{d}y\mathrm{d}\phi} = 2cd_\phi \exp\left\{-2\ln(2d_\phi/\Delta)\zeta + \sqrt{[\ln(2d_\phi/\Delta)\zeta]^2 - y^2}\right\}, \quad (6)$$

where $c$ is a normalization constant. $\zeta$ is a correction parameter representing the corrections for three factors: the initial configuration of the overlap region, the freeze-out condition, and the assumption of an perfect fluid. For example, in calculations, the initial overlap region is taken as a cylinder with thickness $\Delta$, but the reality is that the initial overlap region possesses the shape of an almond being Lorentz contracted along its edge. Furthermore, the freeze-out of the fluid element is supposed to take place as $\rho(t) = d_\phi$. However, the reality may be somewhat different from that [30]. Finally, Eq. (6) is tenable only for a perfect fluid. In realistic case, this equation may have some changes. To take these uncertainties into account, we adopt parameter $\zeta$ to stand for the contributions from them. Since our theoretical knowledge has not advanced to such an extent to determine $\zeta$ in theory, it can now only be fixed by comparing with experimental data.

The total number of the produced charged particles from different azimuthal angles $\phi$ is

$$\frac{\mathrm{d}N_{\text{Fluid}}(y,\sqrt{s})}{\mathrm{d}y} = \int \frac{\mathrm{d}^2 N_{\text{Fluid}}(y,\sqrt{s},\phi)}{\mathrm{d}y\mathrm{d}\phi} \mathrm{d}\phi. \quad (7)$$

It is a function of rapidity and incident energy.

Apart from the charged particles resulted from fluid evolution, leading particles also have contributions to the produced charged particles [31]. In *p-p* collisions, there are only two leading particles. One is in projectile fragmentation region, and the other is in target fragmentation region. Considering that, for a given incident energy, the leading particles resulted in each time of *p-p* collisions have approximately the same amount of energy, then, according to the central limit theorem [32, 33], the leading particles should follow the Gaussian rapidity distribution. That is

$$\frac{\mathrm{d}N_{\text{Lead}}(y,\sqrt{s})}{\mathrm{d}y} = \frac{1}{\sqrt{2\pi}\sigma}\exp\left\{-\frac{[|y| - y_0(\sqrt{s})]^2}{2\sigma^2}\right\}, \quad (8)$$

where $\sigma$ and $y_0(\sqrt{s})$ are the width and central position of Gaussian distribution. In fact, as is known to all, the rapidity distribution of any charged particles produced in hadron or heavy collisions can be well represented by Gaussian form [34−36]. Seeing that, for leading particles resulted in *p-p* collisions at different energies, the relative energy differences among them should not be too much, $\sigma$ should be independent of energy and therefore approximately remain a constant. It is evident that $y_0(\sqrt{s})$ should increase with energy. Both $\sigma$ and $y_0(\sqrt{s})$ can now only be determined by comparing with experimental data.

3. Having the rapidity distribution of Eqs. (7) and (8), the pseudorapidity distribution measured in experiments can be expressed as [37]

$$\frac{dN(\eta,\sqrt{s_{NN}})}{d\eta} = \sqrt{1 - \frac{m^2}{m_T^2 \cosh^2 y}} \frac{dN(y,\sqrt{s_{NN}})}{dy}, \qquad (9)$$

$$y = \frac{1}{2}\ln\left[\frac{\sqrt{p_T^2 \cosh^2 \eta + m^2} + p_T \sinh \eta}{\sqrt{p_T^2 \cosh^2 \eta + m^2} - p_T \sinh \eta}\right], \qquad (10)$$

where $p_T$ is the transverse momentum, $m_T = \sqrt{m^2 + p_T^2}$ is the transverse mass, and

$$\frac{dN(y,\sqrt{s_{NN}})}{dy} = \frac{dN_{Fluid}(y,\sqrt{s_{NN}})}{dy} + \frac{dN_{Lead}(y,\sqrt{s_{NN}})}{dy} \qquad (11)$$

is the total rapidity distribution from both fluid evolution and leading particles.

Experiments have shown that the overwhelming majority of the produced charged particles in hadron or heavy ion collisions at high energy consists of pions, kaons, and protons with proportions of about 83%, 12%, and 5%, respectively [38]. Furthermore, the transverse momentum $p_T$ changes very slowly with beam energies [24, 25]. In calculations, the $p_T$ in Eqs. (9) and (10) takes the values *via* relation [24]

$$p_T = 0.413 - 0.0171\ln(s) + 0.00143\ln^2(s),$$

where $p_T$ and $\sqrt{s}$ are respectively in unit of GeV/c and GeV. The $m$ in Eqs. (9) and (10) takes the values from 0.20 to 0.28 GeV for energies from 23.6 to 900 GeV, which are approximately the mean masses of pions, kaons, and protons.

Substituting Eq. (11) into Eq. (9), we can get the pseudorapidity distributions of the produced charged particles. Figure 1 shows such distributions for *p-p* collisions at $\sqrt{s}$ = 23.6, 45.2, 200, 546, and 900 GeV, respectively. The solid dots are the experimental measurements [20, 21]. The dashed-dotted curves are the results got from the revised Landau hydrodynamic model of Eq. (7). The dotted curves are the results obtained from the leading particles of Eq. (8). The solid curves are the results achieved from Eq. (11), that is, the sums of dashed-dotted and dotted curves. The corresponding $\chi^2/\text{NDF}$ is 0.24, 0.27, 0.23, and 0.27 for $\sqrt{s}$ =23.6, 45.2, 200, and 900 GeV, respectively. It can be seen that the theoretical results are well consistent with experimental measurements.

In calculations, the correction parameter $\zeta$ in Eq. (6) takes the values of 1.17, 1.01, 1.48, 1.69, and 1.74 for energies from small to large. It can be seen that, except for $\sqrt{s}$ =45.2 GeV, $\zeta$ increases with energy. The central parameter $y_0(\sqrt{s})$ in Eq. (8) takes the values of 1.86, 2.23, 2.48, 2.62, and 2.81 for energies from small to large. The $\sigma$ in Eq. (8) takes the value of 0.90 for all concerned incident energies. As the analyses given above, $y_0(\sqrt{s})$ increases with energy, but $\sigma$ remains a constant for different colliding energies.

**4.** Comparing with nucleus-nucleus collisions, *p-p* collisions are the relatively simpler processes. In these processes, the leading particles are better defined and understood. That is, in each time of *p-p* collisions, there are only two leading particles. They are respectively in projectile and target fragmentation region. The charged particles produced in *p-p* collisions are composed of two parts. One is from the hot and dense matter created in collisions, which is supposed to expand and freeze out into measured charged particles in the light of the revised Landau hydrodynamic model. The other is from the leading particles, which are supposed to have a Gaussian rapidity distribution in their respective formation regions.

Comparing with the experimental measurements performed in *p-p* collisions in the whole available energy region from $\sqrt{s}$ =23.6 to 900 GeV, we can get the conclusions as follows:

(1) The revised Landau hydrodynamic model is not only applicable to nuclear interactions but also appropriate to hadronic reactions. This is in accordance with the investigations presented in Ref. [10]. Where, both nucleus-nucleus and *p-p*($\bar{p}$) collisions are assumed to possess the same mechanism of particle production, namely a combination of the constituent quarks in participants together with Landau hydrodynamics. The measurements in nucleus-nucleus reactions are shown to be well reproduced by the measurements in *p-p*($\bar{p}$) interactions.

(2) The centers of the Gaussian rapidity distributions of leading particles increase with

increasing energy. However, the widths of distributions are irrelevant to energy.

(3) It can be seen from Fig. 1 that, though there are only two leading particles in *p-p* collisions, they are essential in describing experimental observations. Only after these two leading particles are taken into account, can the experimental data be matched up properly.

This work is partly supported by the Transformation Project of Science and Technology of Shanghai Baoshan District with Grant No. CXY-2012-25, and by the Shanghai Leading Academic Discipline Project with Grant No. XTKX 2012.

**Figure 1**

The pseudorapidity distributions of the produced charged particles in *p-p* collisions at $\sqrt{s} = 23.6$, 45.2, 200, 546, and 900 GeV, respectively. The solid dots are the experimental measurements [20, 21]. The dashed-dotted curves are the results got from the revised Landau hydrodynamic model of Eq. (7). The dotted curves are the results obtained from leading particles of Eq. (8). The solid curves are the results achieved from Eq. (11), that is, the sums of dashed-dotted and dotted curves.

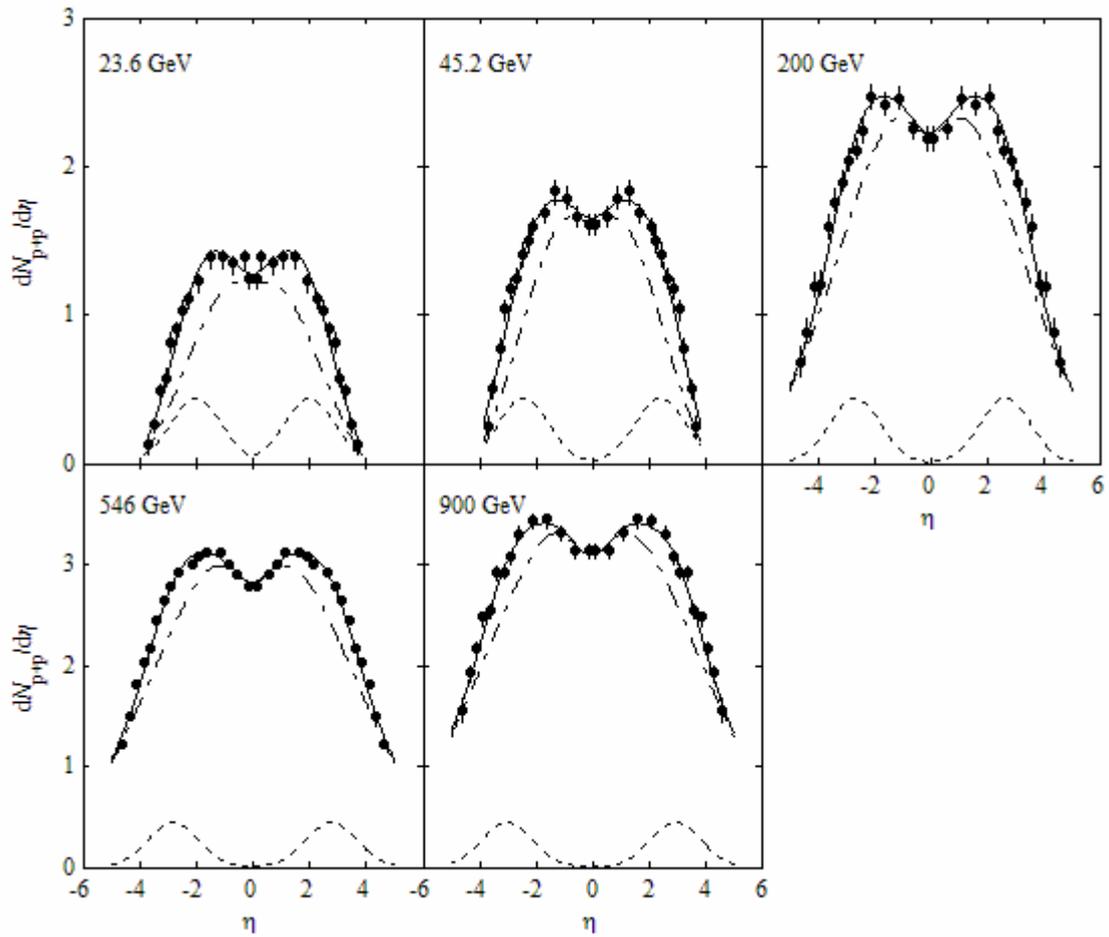

**Figure 1**